\documentclass[%
 reprint,
 amsmath,amssymb,
 aps,
]{revtex4-2}

\usepackage[dvipdfmx]{graphicx} 
\usepackage{dcolumn}
\usepackage{bm}
\usepackage{color} 
\usepackage{braket} 
\usepackage{ulem}

\newcommand{\Slash}[1]{{\ooalign{\hfil/\hfil\crcr$#1$}}} 

\begin{document}

\preprint{}

\title{Modified dilepton production rate from charged pion-pair annihilation in the inhomogeneous chiral condensed phase}

\author{Kentaro Hayashi}
 \email{b22d6a02@s.kochi-u.ac.jp}
\affiliation{Graduate School of Integrated Arts and Sciences, Kochi University, Kochi 780-8520, Japan}

\author{Yasuhiko Tsue}
 \email{tsue@kochi-u.ac.jp}
\affiliation{Department of Mathematics and Physics, Kochi University, Kochi 780-8520, Japan}

\begin{abstract}
We investigate the dilepton production rates from annihilation processes of charged pion pairs 
with modified pion dispersion relations in the inhomogeneous chiral condensed phase.
We assume a dual chiral density wave as an inhomogeneous chiral condensate, and obtain the dispersion relations of the 
Nambu-Goldstone  modes in the inhomogeneous chiral condensed phase. 
We use a low energy effective Lagrangian based on the $O(4)$ symmetry which is expanded by the order parameter 
up to the sixth order.
The obtained dispersion relations are anisotropic and quadratic for the momentum.
We evaluate the electron-positron production rates by charged pion-pair annihilations as functions of an invariant mass 
using the obtained dispersion relations.
Basically, the production rate in the inhomogeneous chiral condensed phase has a steeper overall slope 
with respect to an invariant mass than that in the homogeneous chiral condensed phase.
Therefore, the production rate may be enhanced when the invariant mass is around twice the pion mass.

\end{abstract}

\pacs{Valid PACS appear here}
\maketitle



\section{\label{sec:level1}Introduction}

Elucidating the phase structure of quantum chromodynamics (QCD) is one of the major themes in quark-hadron physics.
In particular, the QCD phase diagram spanned by temperature and density has been much studied 
\ \cite{Fukusima:2011pdd}.
Regarding the high temperature and low density region, a lot of useful information has been obtained from the heavy ion collision experiments and the numerical simulations using the lattice QCD.
In contrast, the low temperature and high density region remains poorly understood because the experimental investigations are more difficult and the lattice simulations currently face technical challenges.
This region has been discussed using low-energy effective theories such as the Nambu-Jona-Lasinio (NJL) model, the quark-meson model, the Ginzburg-Landau approach and so forth.
In this region, it has been pointed out that there may exist the possibilities of various phases such as the color superconductivity phase \cite{Alford:1999cfl, Iida:2002spq, Alford:2008csd}, quarkyonic phase \cite{McLerran:2007pdq}, the spin polarized phase \cite{Nakano:2003spc, Maedan:2007spc, Matsuoka:2016spv}, the inhomogeneous chiral condensed phase \cite{Nakano:2005csa} and so forth.

In recent years, the possibility of various inhomogeneous chiral condensed phases has been theoretically discussed 
\cite{Buballa:2015icc}.
In the conventional QCD phase diagram, it has been thought that a first-order chiral phase transition from hadronic to quark matters occurs at low temperature and high density, although the possibility of a crossover rather than a first-order phase transition has also been pointed out \cite{Hatsuda:2006ncp}.
Studies based on effective theories assuming the existence of inhomogeneous chiral condensate suggest that the region near the chiral transition boundary may be replaced by the inhomogeneous chiral condensed phase \cite{Nakano:2005csa,Nickel:2009ipn}.
It has also been pointed out that the Lifshitz point, where the homogeneous and inhomogeneous chiral condensed phases and chiral restored phase intersect, may coincide with the QCD critical endpoint \cite{Nickel:2009hmp, Carignano:2014ipq}.

In the context of condensed matter physics, inhomogeneous condensates are broadly divided into the Fulde-Ferrell (FF) type \cite{Fulde:1964sss}, characterized by modulations of the phase of a complex order parameter with constant amplitude, and Larkin-Ovchinnikov (LO) type \cite{Larkin:1965nss}, characterized by the modulated amplitude.
In this paper, we adopt the dual chiral density wave (DCDW) \cite{Nakano:2005csa} as a typical inhomogeneous chiral condensate with one-dimensional modulation.
The DCDW consists of a scalar condensate and a pseudoscalar-isovector condensate with third isospin component and is modulated spatially in one dimension with a constant amplitude $\Delta$ and a wave number vector $\bm{q}$, 
\begin{align}
\braket{\bar{\psi}\psi} = \Delta\cos(\bm{q}\cdot\bm{r}), \ 
\braket{\bar{\psi}i\gamma_5\tau_3\psi} = \Delta\sin(\bm{q}\cdot\bm{r}),
\end{align}
where $\psi$ is a two-flavor quark field and $\tau_3$ is a Pauli matrix in the isospin space.
Note that in Ref.\ \cite{Lee:2015lpi}, the DCDW type inhomogeneous chiral condensed phase may be a quasi-long-range order rather than a long-range order, i.e., it may not exist as a stable state at finite temperature. However, even if it is a metastable state, it is expected to affect phenomena in heavy ion collisions 
\cite{TVM} or under strong magnetic fields. 

Recently, experimental projects have been planned to probe the low temperature and high density region of the QCD 
phase diagram {\cite{SAKO}}.
Depending on the phase structure, significant changes in observable quantities due to phase transitions or phase boundaries may be observed, so it is important to discuss possible changes in advance qualitatively or quantitatively.
For example, reference \cite{Rennecke:2023pim} suggests that particle interferometry can be a sensitive probe of  phases with spatially modulations, such as inhomogeneous chiral condensed phase. 
In such phase, particles may have a moat spectrum that has a minimum at nonzero momentum \cite{PR}. 
In Ref.\ \cite{Rennecke:2023pim}, it has been pointed out that such dispersion can cause a change in the structure of the correlation function.
We focus on the dilepton production rate as an observable that varies accompanied by a modification of dispersion relations.
This is because the dilepton production rate from pion-pair annihilation depends on the charged pion dispersion relation \cite{Gale:1987drf}, and lepton emission has the advantage of  the immunity to the strong interaction in the collision region.
Therefore, in this paper we discuss the change in the dilepton production rate due to the charged pion dispersion relation of the inhomogeneous chiral condensed phase.
The dispersion relations for the Nambu-Goldstone (NG) modes in the inhomogeneous chiral condensed phase are derived along the lines of Ref.\ \cite{Lee:2015lpi} using the Ginzburg-Landau approach, and the dilepton production rates are estimated 
assuming a vector meson dominance \cite{Gale:1987drf}.
Note that the Ginzburg-Landau approach is only valid in the region of small wave number and amplitude, such as near the Lifshitz point, and is also valid at low energy.

This paper is organized as follows: 
In the next section, we construct a low energy effective Lagrangian density expanded with respect to the order parameter based on the 
$O(4)$ symmetry, and assume a DCDW as an inhomogeneous chiral condensate.
In Sec.\ \ref{sec:level3}, we derive the dispersion relations of the NG modes in the inhomogeneous chiral condensed phase from the Lagrangian constructed in Sec.\ \ref{sec:level2}.
In Sec.\ \ref{sec:level4}, 
we formulate the dilepton production rate from the pair annihilation of charged pions with anisotropic dispersions.
In Sec.\ \ref{sec:level5}, we evaluate the electron-positron pair production rate from the charged pion pair annihilation in the inhomogeneous chiral condensed phase using the obtained dispersions in Sec.\ \ref{sec:level3}.
The last section is devoted to a summary.


\section{\label{sec:level2}Ginzburg-Landau analysis}

In the previous paper \cite{Lee:2015lpi}, the dispersion relations of the NG modes in the inhomogeneous chiral condensed phase of the DCDW type was derived using the Ginzburg-Landau approach.
We revisit the formulation of that approach in this section, and re-derive the dispersion relations under slightly different conditions in the next section.

\subsection{\label{sec:level2-1}Lagrangian}

We consider a system with the chiral $SU(2)_{L} \times SU(2)_{R}$ symmetry. Since $SU(2)_{L} \times SU(2)_{R}$ chiral symmetry is isomorphic to four-dimensional rotation group $O(4)$, we introduce four-component field $\phi^{T} = (\sigma, \vec{\pi})$ as an order parameter in the context of the Ginzburg-Landau approach. 
This field $\phi$ transforms as a four-dimensional vector under $O(4)$ rotations:
$U(\alpha_i) = \exp[-i\alpha_i L_{Vi}]\ (i=1,2,3)$ and $U(\beta_j) = \exp[-i\beta_j L_{Aj}]\ (j=1,2,3)$, where $L_{V1,2,3}$ and $L_{A1,2,3}$ are $O(4)$ generators and $\alpha_{1,2,3}$ and $\beta_{1,2,3}$ are the rotation angles.
For $\phi$, $U(\alpha_{1,2,3})$ and $U(\beta_{1,2,3})$ represent the $\pi_2$-$\pi_3, \pi_3$-$\pi_1, \pi_1$-$\pi_2, \sigma$-$\pi_1, \sigma$-$\pi_2$ and $\sigma$-$\pi_3$ rotations, respectively.
In addition, $\sigma$ and $\vec{\pi}$ are related to chiral scalar condensate $\braket{\bar{\psi}\psi}$ and 
chiral pseudoscalar-isovector condensate $\braket{\bar{\psi}i\gamma_5\vec{\tau}\psi}$, respectively.
Therefore, the rotations with $\alpha_{1,2,3}$ correspond to the vector (isospin) rotations while those with $\beta_{1,2,3}$ correspond to the axial vector (isospin) rotations.
The correspondence between the chiral $SU(2)_{L} \times SU(2)_{R}$  symmetry and $O(4)$ symmetry is discussed in Appendix \ref{app:SU2_O4}.

Now, we start by considering a low energy effective Lagrangian density in terms of $\phi$ and its derivatives,
\begin{align}
\mathcal{L} &= \tilde{c}|\partial_t \phi|^2 -V(\phi) ,
\label{eq:L_GL} \\
\notag \\
V(\phi) &=a|\phi|^2 +b|\phi|^4 +c|\nabla\phi|^2 +d|\nabla^2\phi|^2 \notag \\
            &\quad +e|\nabla\phi|^2 \cdot|\phi|^2 +f|\phi|^6 +g(\phi\cdot\nabla\phi)^2 .
\label{eq:V_GL}	
\end{align}
We assume that density effects are implicitly subsumed in the coefficients $a\sim g$ and Lorentz invariance is explicitly broken. 
The potential term $V(\phi)$ is expanded up to sixth order in powers of the field and fourth order in its derivatives, as required for stability of the inhomogeneous phase at the mean-field level.
Note that the following discussion is limited to low-energy fluctuations since we ignore higher order terms of $\mathcal{O}(\phi^8)$ and $\mathcal{O}(\nabla^6)$.
In the following, we set {$\tilde{c}=1$} for simplicity and regard the expansion coefficients, which depend the density and the temperature, as free parameters.

\subsection{\label{sec:level2-2}Dual chiral density wave}

As the order parameter of the inhomogeneous chiral condensed phase, we adopt a time independent and spatial inhomogeneous field of the DCDW type
\begin{align}
\phi_0 = \Delta 
\begin{pmatrix}
\cos (\bm{q}\cdot\bm{r}) \\
0 \\
0 \\
\sin (\bm{q}\cdot\bm{r})
\end{pmatrix}
= \Delta e^{-i\bm{q}\cdot\bm{r}L_{A3}}
\begin{pmatrix}
1 \\
0 \\
0 \\
0
\end{pmatrix} ,
\label{eq:DCDW_gs}
\end{align}
where $\Delta$ is a constant amplitude, $\bm{q}$ is the wave number vector of the spatial modulation and $\bm{r} \equiv (x,y,z)$.
By substituting $\phi=\phi_0$ to Eq.~\eqref{eq:V_GL}, the potential term of the Lagrangian density becomes
\begin{align}
V(\phi_0) = a\Delta^2 +b\Delta^4 +cq^2\Delta^2 +dq^4\Delta^2 +eq^2\Delta^4 +f\Delta^6 ,
\label{eq:V_phi_0}	
\end{align}
where $q=|\bm{q}|$.
From the stability we obtain the conditions for the coefficients: $f>0$, $d>0$ and $df - e^2/4 >0$.
Considering $\phi_0$ as the ground state, the values of $\Delta$ and $q$ are determined by minimizing Eq.~\eqref{eq:V_phi_0}.
Then, we can get the following gap equations for $\Delta$ and $q$ from the stationary conditions ${\partial V}/{\partial q}={\partial V}/{\partial \Delta}=0$: 
\begin{align}
q\Delta^2(c +2dq^2 +e\Delta^2) =0 ,
\label{eq:gap1}
\end{align}
\begin{align}
\Delta[a +2b\Delta^2 +(c+2e\Delta^2)q^2 +dq^4 +3f\Delta^4] =0 .
\label{eq:gap2}
\end{align}
There are three types of solutions for $\Delta$ and $q$:
(i)$\Delta=0$, namely, chiral symmetry restored phase,
(ii)$\Delta\ne0, q=0$, namely, homogeneous chiral condensed phase,
(iii)$\Delta\ne0, q\ne0$, namely, inhomogeneous chiral condensed phase.

Since the coefficients depend on temperature and density, the values of $\Delta$ and $q$ are also determined by them.
Therefore, by explicitly defining the coefficients as functions of thermodynamic variables such as temperature and chemical potentials, the phase structure for those parameters can be obtained at the classical level.
Although we treat the coefficients as free parameters throughout this paper, we assume that the inhomogeneous chiral condensed phase is realized for a given set of coefficients. 

\subsection{\label{sec:level2-3}Symmetry breaking and NG modes}

The inhomogeneous chiral condensate of the DCDW type, Eq. \eqref{eq:DCDW_gs}, with nonvanishing $\Delta$ and $q$ 
breaks space symmetries and the chiral $SU(2)_{L} \times SU(2)_{R}$ symmetry 
spontaneously, namely $O(4)$ symmetry.
When wave number vector of modulation $\bm{q}$ is parallel to the $z$ axis, the broken symmetries are five symmetries of the $O(4)$  symmetry which correspond to $U(\alpha_{1,2}), U(\beta_{1,2,3})$, the translational symmetry in the $z$ direction and the symmetries under rotations about the $x$ and $y$ axes.
However, these broken generators are not independent.
Here we set the following parameters: the chiral rotation angles $\vec{\alpha}, \vec{\beta}$,  translational displacement of the $z$ direction $s_z$ and the angles under rotations about the $x$ and $y$ axes $\theta_x, \theta_y$.
To see the symmetry breaking pattern explicitly, we perform infinitesimal transformations on Eq. \eqref{eq:DCDW_gs}:
\begin{align}
\phi_0 
\to \phi_0 + \Delta
\begin{pmatrix}
-[q(s_z +y\theta_x -x\theta_y) +\beta_3]\sin qz \\
\beta_1\cos qz +\alpha_2\sin qz \\
\beta_2\cos qz -\alpha_1\sin qz \\
 [q(s_z +y\theta_x -x\theta_y) +\beta_3]\cos qz 
\end{pmatrix} .
\end{align}

The form of the first and the fourth components implies that $s_z, y\theta_x, x\theta_y$ and $\beta_3$ yield the same transformation for $\phi_0$.
In terms of symmetry generators, one of the linear combinations of four generators corresponding to $s_z, y\theta_x, x\theta_y$ and $\beta_3$ is the broken generator corresponding to the NG mode. 
Similar arguments of a coupling of translational and internal symmetry and a coupling of translational and rotational symmetry are given in Refs.\ \cite{Kobayashi:2014nrn} and \cite{Low:2002sbs}.

Similarly, the second component shows that $\beta_1$ and $\alpha_2$ are not independent in the sense discussed in Refs.\ \cite{Low:2002sbs, Watanabe:2013rng, Hayata:2014bss}. 
Therefore, one of the linear combinations of two generators corresponding to $\beta_1$ and $\alpha_2$ is the broken generator corresponding to the NG mode. 
Similar arguments can be applied to $\beta_2$ and $\alpha_1$ from the third component.

Finally, there are only three independent NG modes: $q(s_z +y\theta_x -x\theta_y) +\beta_3$, $\beta_1\cos qz +\alpha_2\sin qz$ and $\beta_2\cos qz -\alpha_1\sin qz$. 
In the following, we set the replacement $q(s_z +y\theta_x -x\theta_y) +\beta_3 \to \beta_3$ for simplicity and ignore $s_z, y\theta_x$ and $x\theta_y$.
In contrast, although $\alpha_1=\alpha_2=0$ is adopted in the previous paper \cite{Lee:2015lpi}, 
$\beta_1$, $\beta_2$, $\alpha_1$ and $\alpha_2$ should remain in this paper because each of them is accompanied by a modulation effect ($\sin qz$ or $\cos qz$).


\section{\label{sec:level3}Dispersion relations of NG modes in DCDW phase}

We now consider low energy excitations in inhomogeneous chiral condensed phase of DCDW type.
We add the amplitude fluctuation $\delta$ and the phase fluctuations $\beta_1, \beta_2, \beta_3, \alpha_1, \alpha_2$ ,which correspond to the broken generators, to the ground state \eqref{eq:DCDW_gs}:
\begin{align}
\phi
&= (\Delta +\delta)  e^{-i\alpha_1L_{V1}}  e^{-i\alpha_2L_{V2}}   \notag \\
&\quad \times e^{-i\beta_1L_{A1}} e^{-i\beta_2L_{A2}} e^{-i\beta_3L_{A3}}
\begin{pmatrix}
\cos (\bm{q}\cdot\bm{r}) \\
0 \\
0 \\
\sin (\bm{q}\cdot\bm{r})
\end{pmatrix} .
\label{eq:phi_fluc}
\end{align}
Here $L_{A1,A2,A3}$ are $O(4)$ generators of axial isospin rotations and $L_{V1,2,3}$ are $O(4)$ generators of isospin rotations.
Taking up to the first order of fluctuations, Eq. \eqref{eq:phi_fluc} yields
\begin{align}
\phi 
= \phi_0 + \Delta
\begin{pmatrix}
\frac{\delta}{\Delta}\cos(\bm{q}\cdot\bm{r}) -\beta_3\sin(\bm{q}\cdot\bm{r}) \\
\beta_1\cos(\bm{q}\cdot\bm{r}) +\alpha_2\sin(\bm{q}\cdot\bm{r}) \\
\beta_2\cos(\bm{q}\cdot\bm{r}) -\alpha_1\sin(\bm{q}\cdot\bm{r}) \\
\frac{\delta}{\Delta}\sin(\bm{q}\cdot\bm{r})  + \beta_3\cos(\bm{q}\cdot\bm{r})
\end{pmatrix} .
\end{align}

In this paper, we consider local fluctuations $\delta(x)$, $\alpha_{1,2}(x)$ and ${\beta_{1,2,3}}(x)$, where $x \equiv (t, \bm{r})$.
Here we use the Euler-Lagrange equation to obtain the behaviors of the fluctuations.
From the principle of least action, the Euler-Lagrange equations for the Lagrangian density \eqref{eq:L_GL} with 
\eqref{eq:V_GL} are given as 
\begin{align}
\frac{\partial \mathcal{L}}{\partial \phi} -\partial_{\mu}\frac{\partial \mathcal{L}}{\partial (\partial_{\mu}\phi)}
+\nabla^2 \frac{\partial \mathcal{L}}{\partial(\nabla^2 \phi)} &=0 .
\label{eq:EL}
\end{align}
Substituting the Lagrangian density \eqref{eq:L_GL} with \eqref{eq:V_GL} for Eq. \eqref{eq:EL}, we get the formula
\begin{align}
&-\partial_t^2 \phi - [a +2b|\phi|^2 +e|\nabla\phi|^2 +3f|\phi|^4 -g\nabla(\phi\cdot\nabla\phi)]\phi  \notag \\
& \quad +(c +e|\phi|^2)\nabla^2\phi +e\nabla\phi \cdot \nabla|\phi|^2-d\nabla^4\phi =0 .
\label{eq:EL_phi}
\end{align}
Plugging Eq. \eqref{eq:phi_fluc} into Eq. \eqref{eq:EL_phi} and taking up to the first order of fluctuations, we get the following equations for the fields of fluctuations: 
\begin{widetext}
\begin{align}
\begin{pmatrix}
\partial_t^2  -4d(\bm{q}\cdot\nabla)^2 +d\nabla^4 & 2(2d\nabla^2 -e\Delta^2)(\bm{q}\cdot\nabla) \\
-2(2d\nabla^2 -e\Delta^2)(\bm{q}\cdot\nabla) & \partial_t^2 -4d(\bm{q}\cdot\nabla)^2 -g\Delta^2\nabla^2 +d\nabla^4 +4\Delta^2(b +eq^2 +3f\Delta^2) \\
\end{pmatrix}
\begin{pmatrix}
\Delta\beta_3 \\
\delta
\end{pmatrix}
= 0 ,
\label{eq:Matrix_delta_beta3}
\end{align}
\begin{align}
\begin{pmatrix}
\partial_t^2 +d\nabla^4 -4d(\bm{q}\cdot\nabla)^2 & 4d\bm{q}\cdot\nabla^3 \\
-4d\bm{q}\cdot\nabla^3 & \partial_t^2 +d\nabla^4 -4d(\bm{q}\cdot\nabla)^2 \\
\end{pmatrix}
\begin{pmatrix}
\Delta\beta_1 \\
\Delta\alpha_2
\end{pmatrix}
= 0 .
\label{eq:Matrix_beta1_alpha2}
\end{align}
\end{widetext}
Moreover, the equation similar to Eq. \eqref{eq:Matrix_beta1_alpha2} also exists for $\beta_2$ and $\alpha_1$. In these equations, the gap equations \eqref{eq:gap1} and \eqref{eq:gap2} have been used. The matrices in Eqs. \eqref{eq:Matrix_delta_beta3} and  \eqref{eq:Matrix_beta1_alpha2} have nonvanishing off-diagonal elements, so there are the $\delta$-$\beta_3$, $\beta_1$-$\alpha_2$ and $\beta_2$-$\alpha_1$ mixings. Three NG modes should correspond to these mixings respectively. Therefore, the dispersion relations of the NG modes can be obtained by diagonalizing the matrices in Eqs. \eqref{eq:Matrix_delta_beta3} and  \eqref{eq:Matrix_beta1_alpha2}.

Now we replace the space-time differential with energy-momentum:$(\partial_t, \nabla) \to (-i\omega, +i\bm{k})$. 
By assuming that the determinant of the matrix in Eq. \eqref{eq:Matrix_delta_beta3} is equal to zero, we obtain the dispersion relations of the modes for the $\delta$-$\beta_3$ mixing,
\begin{align}
\omega_{\pm}^2 &= 4d(\bm{q}\cdot\bm{k})^2+\frac{1}{2}g\Delta^2k^2 +dk^4 +\frac{1}{2}{m_0}^2 \notag \\
           &\quad \pm \sqrt{ \left(\frac{{m_0}^2}{2} +\frac{g\Delta^2k^2}{2} \right)^2 +4(2dk^2 +e\Delta^2)^2(\bm{q}\cdot\bm{k})^2} ,
\label{eq:dispersion_delta_beta3_1}
\end{align}
where ${m_0}^2 \equiv 4(b+eq^2+3f\Delta^2)\Delta^2$ and $k=|\bm{k}|$.
For small $k$, Eq. \eqref{eq:dispersion_delta_beta3_1} is expanded as
\begin{align}
\omega_{+}^2 
&={m_0}^2 +4dq^2 \left( 1+ \frac{e^2\Delta^4}{d{m_0}^2} \right) k^2 \cos^2\theta +g\Delta^2k^2  \notag \\
&\quad + Ak^4 \cos^2\theta -Bk^4 \cos^4\theta +dk^4 ,
\label{eq:dispersion_delta_beta3_2}
\\ \notag \\
\omega_{-}^2 
&= 4dq^2 \left( 1- \frac{e^2\Delta^2}{4ds} \right)k^2 \cos^2\theta  \notag \\
&\quad -Ak^4\cos^2\theta +Bk^4 \cos^4\theta +dk^4 ,
\label{eq:dispersion_delta_beta3_3}
\end{align}
where 
\begin{align}
A = 4e\Delta^2q^2 \left( \frac{4d}{{m_0}^2}-\frac{eg\Delta^4}{{m_0}^4} \right) ,\ B = \frac{16e^4\Delta^8 q^4}{{m_0}^6} ,
\end{align}  
and $\theta$ is the angle between $\bm{q}$ and $\bm{k}$: $\bm{q}\cdot\bm{k}=qk\cos\theta$.
These dispersion relations have anisotropy and 
$\omega_{-}$ in \eqref{eq:dispersion_delta_beta3_3} corresponds to the NG (massless) mode.

Similarly, by assuming that the determinant of the matrix in Eq. \eqref{eq:Matrix_beta1_alpha2} is equal to zero, we obtain the dispersion relations of the modes for the $\beta_1$-$\alpha_2$ mixing,
\begin{align}
\omega_i^2 = 4dq^2k^2 \cos^2\theta \pm 4d q k^3 \cos\theta +dk^4
\label{eq:dispersion_beta1_alpha2}
\end{align}
The same formula is derived for $\beta_2$-$\alpha_1$ mixing.

We would like to regard Eq. \eqref{eq:dispersion_beta1_alpha2} as the dispersion of the charged pion since the modes of the $\beta_1$-$\alpha_2$ and $\beta_2$-$\alpha_1$ mixings correspond to isospin one-component and two-component pions, respectively 
(See, Appendix \ref{app:Mixed_Mode}). 
However, Eq. \eqref{eq:dispersion_beta1_alpha2} has two problems: (i) It is not invariant under momentum reversal $\bm{k} \to -\bm{k}$. (ii)There are two massless dispersions for one NG mode.
These problems seem to arise from the fact that the wave number vector of the modulation $\bm{q}$ formally has a positive and negative direction.
If $\bm{k}\to-\bm{k}$ in one equation of Eq. \eqref{eq:dispersion_beta1_alpha2}, it becomes equal to the other equation.
In other words, one equation of Eq. \eqref{eq:dispersion_beta1_alpha2} for $\theta$ is the same as the other equation for $\theta+\pi$.
Therefore, in this paper, we allow the spatial inversion asymmetry of each dispersion and add up their contributions.
In addition, we add the bare charged pion mass $m_{\pi}$ to each dispersion.
\begin{align}
\omega_i^2 = m_{\pi}^2 + 4dq^2k^2 \cos^2\theta \pm 4dq k^3 |\cos\theta| +dk^4 .
\label{eq:dispersion_chargedpion}
\end{align}
Of course, $m_{\pi}$ should be derived from the explicit $O(4)$ symmetry breaking term which corresponds to the effect of the nonvanishing quark current mass.

\begin{figure}[h]
  \begin{minipage}[b]{1\linewidth}
    \includegraphics[keepaspectratio, height=4.5cm]{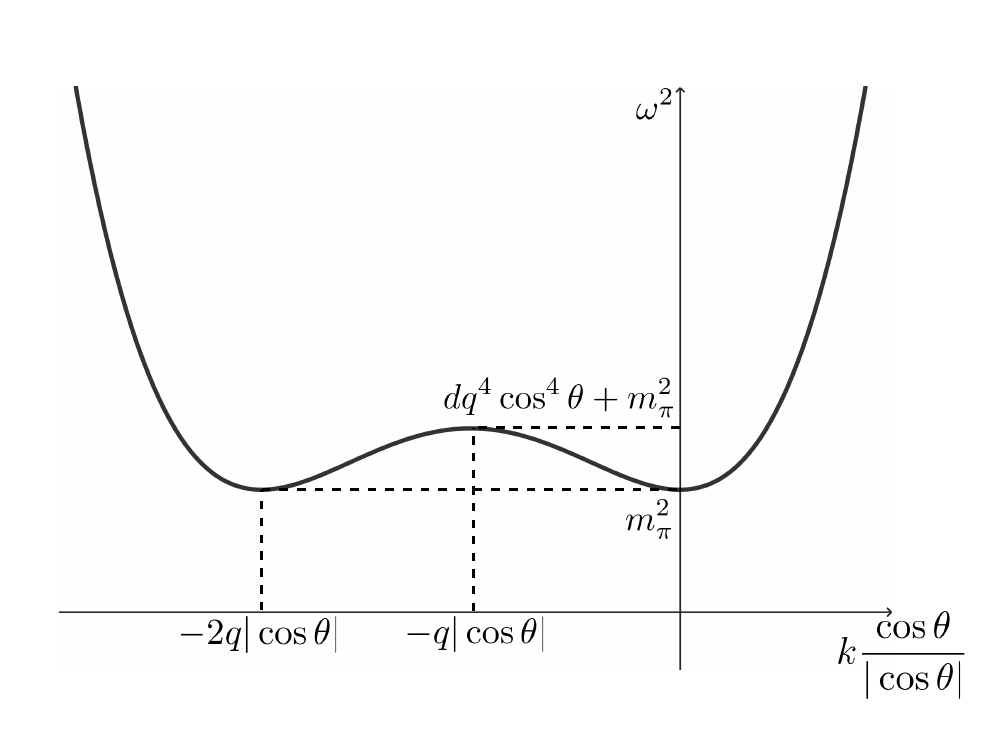}
  \end{minipage}
\\
  \begin{minipage}[b]{1\linewidth}
    \includegraphics[keepaspectratio, height=4.5cm]{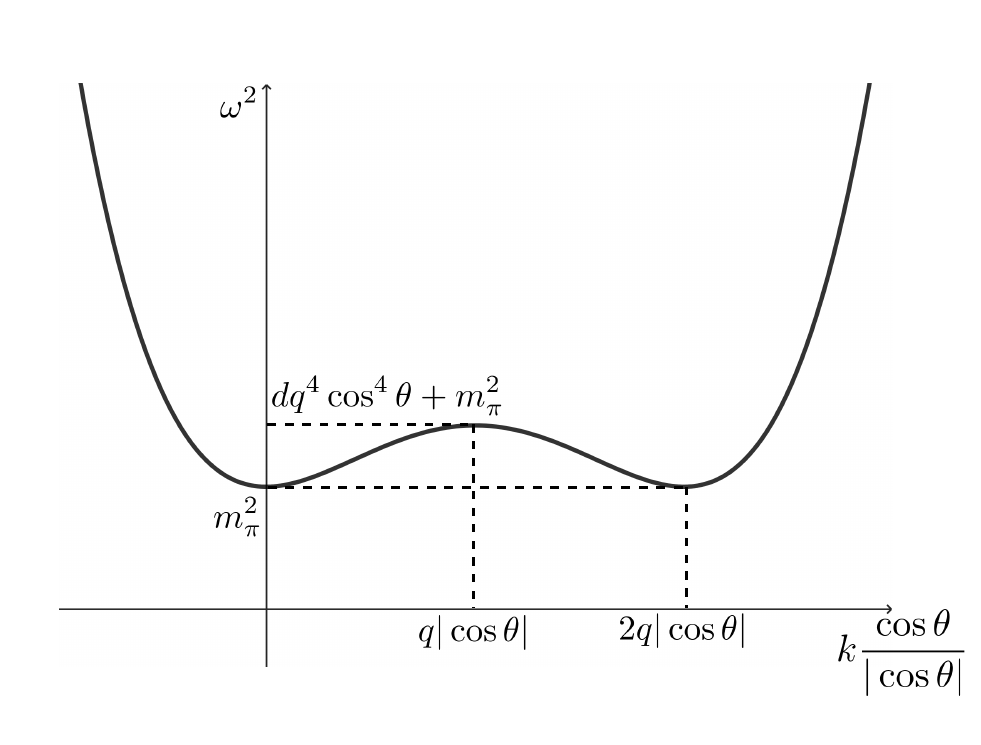}
  \end{minipage}
  \caption{The upper (lower) panel is the sketch of the graph of $\omega^2=f_+(k)+m_{\pi}^2$ ($\omega^2=f_-(k)+m_{\pi}^2$) for a certain value of $\theta$, with $\omega^2$ on the vertical axis and $k\cos\theta/|\cos\theta|$ on the horizontal axis.}
\label{fig:dispersion_f+_f-}
\end{figure}

Now, we consider each mode of dispersion and their relationship. 
We can expand $\beta_1(x)$ and $\alpha_2(x)$ as follows:
\begin{align}
\beta_1(x) &=\int d^4k \bar{\beta}_1(k) e^{-ikx} ,
\label{eq:expansion_beta1} \\
\alpha_2(x) &=\int d^4k \bar{\alpha}_2(k) e^{-ikx} ,
\label{eq:expansion_alpha2}
\end{align}
where $kx=\omega t - \bm{k}\cdot\bm{x}$.
By substituting these expansions in Eq. \eqref{eq:Matrix_beta1_alpha2} and diagonalizing the matrix in that equation, we obtain
\begin{align}
\begin{pmatrix}
\omega^2 -f_+(\bm{k}) & 0 \\
0 & \omega^2 -f_-(\bm{k})
\end{pmatrix}
\begin{pmatrix}
\Delta \left( \bar{\beta}_1(k) +i\bar{\alpha}_2(k) \right) \\
\Delta \left( \bar{\beta}_1(k) -i\bar{\alpha}_2(k) \right)
\end{pmatrix}
= 0 ,
\label{eq:Matrix_beta1_alpha2_k1}
\end{align}
where
\begin{align}
f_{\pm}(\bm{k}) 
=d(\bm{k}^2)^2 \pm 4d(\bm{q}\cdot\bm{k})\bm{k}^2 +4d(\bm{q}\cdot\bm{k})^2\ .
\end{align}
Of course, by adding the pion mass by hand, we can also get Eq.\ \eqref{eq:dispersion_chargedpion} from \eqref{eq:Matrix_beta1_alpha2_k1}.
Figure \ref{fig:dispersion_f+_f-} is the sketch of the functional form of the dispersion relations \eqref{eq:dispersion_chargedpion} for a certain value of $\theta$.
Since $f_+(k)$ and $f_-(k)$ have the following properties: $f_+(\bm{k}-2\bm{q})=f_-(\bm{k}),\ f_-(\bm{k}+2\bm{q})=f_+(\bm{k}),\ f_+(-\bm{k}-2\bm{q})=f_+(\bm{k}),\ f_-(-\bm{k}+2\bm{q})=f_-(\bm{k})$, Eq.\ \eqref{eq:Matrix_beta1_alpha2_k1} can be rewritten as
\begin{align}
\begin{pmatrix}
\omega^2 -f_+(\bm{k}) & 0 \\
0 & \omega^2 -f_-(\bm{k})
\end{pmatrix}
\begin{pmatrix}
\frac{1}{4}\Delta \bar{\gamma}_{+}(k) \\
\frac{1}{4}\Delta \bar{\gamma}_{-}(k)
\end{pmatrix}
= 0 ,
\label{eq:Matrix_beta1_alpha2_k2}
\end{align}
where
\begin{align}
&\bar{\gamma}_{+}(k) = \bar{\beta}_1(k) +i\bar{\alpha}_2(k) +\bar{\beta}_1(k+2\bm{q}) -i\bar{\alpha}_2(k+2\bm{q}) \notag \\
                &\quad +\bar{\beta}_1(-k-2\bm{q}) +i\bar{\alpha}_2(-k-2\bm{q}) +\bar{\beta}_1(-k) -i\bar{\alpha}_2(-k) , 
\label{eq:gamma+} \\
&\bar{\gamma}_{-}(k) = \bar{\beta}_1(k) -i\bar{\alpha}_2(k) +\bar{\beta}_1(k-2\bm{q}) +i\bar{\alpha}_2(k-2\bm{q}) \notag \\
                &\quad +\bar{\beta}_1(-k+2\bm{q}) -i\bar{\alpha}_2(-k+2\bm{q}) +\bar{\beta}_1(-k) +i\bar{\alpha}_2(-k) .
\label{eq:gamma-}
\end{align}

\begin{figure}[h]
\centering
\includegraphics{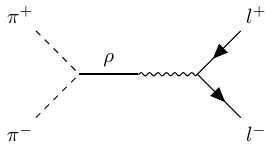}
\caption{$\pi^{+} \pi^{-} \to \rho \to \gamma^{*} \to l^{+} l^{-}$. By the vector meson dominance, the vertex of pion pair annihilation and virtual photon production can be approximated to a process through a rho meson which subsequently decays into a virtual photon. }
\label{fig:feynman_rho}
\end{figure}

\section{\label{sec:level4}dilepton production rate}

We consider the dilepton production by charged pion-pair annihilation: $\pi^{+} \pi^{-} \to l^{+} l^{-}$.
In this paper we assume the vector meson dominance \cite{VMD}.
Therefore, at zero density, this process can be approximated to the process through the rho meson in Fig.\ \ref{fig:feynman_rho}: $\pi^{+} \pi^{-} \to \rho \to \gamma^{*} \to l^{+} l^{-}$.
However, the inhomogeneous chiral condensed phase can only be realized in a finite density region. 
Therefore, we might need to include the corrections of $\Delta$ excitations by $\pi$-$N$ coupling in that region \cite{Korpa:1990wlp}.
However, in this paper, we focus on the process in Fig.\ \ref{fig:feynman_rho} because we would like to consider the effects of the charged pion dispersion relation on the lepton pair production rates in the inhomogeneous chiral condensed phase.

Now, we set the energy-momentum of each particle in the process of Fig.\ \ref{fig:feynman_rho} as follows:
$\pi_{+}:(\omega_1,\bm{k}_1),\ \pi_{-}:(\omega_2,\bm{k}_2),\ l^{\pm}:(E_{\pm},\bm{p}_{\pm}),\ \gamma^{*}:(M,\bm{{Q}})$.
The virtual photon's energy $M$ represents the invariant mass of the system.
Here we consider the dilepton production rate for the invariant mass {$M$} in the center of mass system, 
namely $\bm{{Q}}=0$:

\begin{widetext}
\begin{align}
\left. \frac{d^4 R}{d^3{Q} dM} \right|_{\bm{Q} = 0}
&= \int\frac{d^3k_1}{2\omega_1(2\pi)^3} \frac{1}{e^{\omega_1 \beta} -1} 
   \int\frac{d^3k_2}{2\omega_2(2\pi)^3} \frac{1}{e^{\omega_2 \beta} -1} 
   \int\frac{d^3p_{-}}{2E_{-}(2\pi)^3} \int\frac{d^3p_{+}}{2E_{+}(2\pi)^3} |F_{\pi}(M^2)|^2 \notag \\
   &\quad \times |\mathcal{M}|^2
          (2\pi)^4 \delta(\omega_1 +\omega_2 -E_{-} -E_{+}) 
           \delta(\bm{k}_1 +\bm{k}_2 -\bm{p}_{-} -\bm{p}_{+})
          \delta(\bm{k}_1 +\bm{k}_2) \delta(M -\omega_1 -\omega_2)\ ,
\label{eq:PR_1}
\end{align}
\end{widetext}
where $\beta$ is the inverse of temperature, namely $\beta=1/T$,  $\mathcal{M}$ is the 
invariant scattering amplitude, 
\begin{align}
i\mathcal{M} 
\sim
\vcenter{\hbox{\includegraphics[clip,scale=0.8]{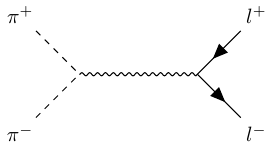}}}\ ,
\label{eq:iM}
\end{align}
and $F_{\pi}(M^2)$ is the form factor by rho meson dominance.
\begin{align}
F_{\pi}({Q}^2) = \frac{m_{\rho}^2}{{Q}^2 -m_{\rho}^2 +i\Gamma_{\rho}m_{\rho}} \ .
\label{eq:form_factor}
\end{align}
Here, $m_{\rho}$ is the rho meson mass and $\Gamma_{\rho}$ is the decay width of rho meson.
The case which the dispersion of charged pions is isotropic has already been formulated in Ref.\ \cite{Gale:1987drf}.
We extend it to the anisotropic case since the dispersion relations obtained in the previous section have an angular dependence to the modulation wave number vector.
So let us set $\omega_i=\omega_i(k_i,\theta_i)$.
Furthermore, $\omega_i(k_i,\theta_i+\pi)=\omega_i(k_i,\theta_i)$ for a dispersion with spatial inversion symmetry such as Eq.\ \eqref{eq:dispersion_chargedpion}.
After tedious calculations (See Appendix \ref{app:Formulation_PR}), Eq. \eqref{eq:PR_1} yields

\begin{align}
\left. \frac{d^4 R}{d^3{Q} dM} \right|_{\bm{{Q}} = 0}
= \int_0^{\pi} \left. \frac{d^5 R}{d^3{Q} dMd\theta} \right|_{\bm{{Q}} = 0} d\theta \ ,
\label{eq:integration_theta}
\end{align}
\begin{align}
\left. \frac{d^5 R}{d^3{Q} dMd\theta} \right|_{\bm{{Q}} = 0}
&= \sum_{j} \frac{8\alpha^2}{3(2\pi)^4} \frac{M^2 +2m_l^2}{M^6} |F_{\pi}(M^2)|^2 \notag \\
   &\quad \times \frac{k^{(j)4} \sin\theta}{(e^{\beta M/2}-1)^2} 
      \left| \frac{d\bar{\omega}}{dk} \right|_{k=k^{(j)}, 2\bar{\omega} = M}^{-1} \ ,
\label{eq:produtionrate_theta}
\end{align}
where $\alpha$ is the fine-structure constant ($\sim 1/137$), $m_l$ is a lepton mass, $\bar{\omega} \equiv ({\omega_1+\omega_2})/{2} = \omega_1=\omega_2$, and $k^{(j)}$ is the solution of $k$ that satisfies $2\bar{\omega}(k,\theta) =M$ for a certain $\theta$.


\section{\label{sec:level5}Numerical results}

By substituting Eq. \eqref{eq:dispersion_chargedpion} for $\bar{\omega}$ in Eq. \eqref{eq:produtionrate_theta}, we obtain
\begin{align}
\left. \frac{d^5 R}{d^3{Q} dMd\theta} \right|_{\bm{{Q}} = 0} 
&= \sum_{j} \frac{\alpha^2}{24\pi^4} \frac{M^2 +2m_l^2}{M^5} \frac{|F_{\pi}(M^2)|^2}{(e^{\beta M/2}-1)^2} \notag \\
   &\times \frac{k_1^{(j)3} \sin\theta}{|2dq^2\cos^2\theta \pm 3dqk^{(j)} \cos\theta +dk^{(j)2} |}.
\label{eq:produtionrate_cal}
\end{align}
By performing the integration of Eq.\ \eqref{eq:integration_theta} and adding together the contributions from the cases of $f_+(\bm{k})$ and $f_-(\bm{k})$, we can calculate the dilepton production rate.

In this section, we show the numerical results of the electron-positron pair production rates.
Incidentally, we can also calculate the muon pair production rates, which is $(M^2 +2m_{\mu}^2)/(M^2 +2m_{e}^2)$ times larger than the electron-positron pair production rates for electron mass $m_{e}$ and muon mass $m_{\mu}$.


\begin{figure*}[tbp]
 \centering
 \includegraphics[keepaspectratio, height=5cm]{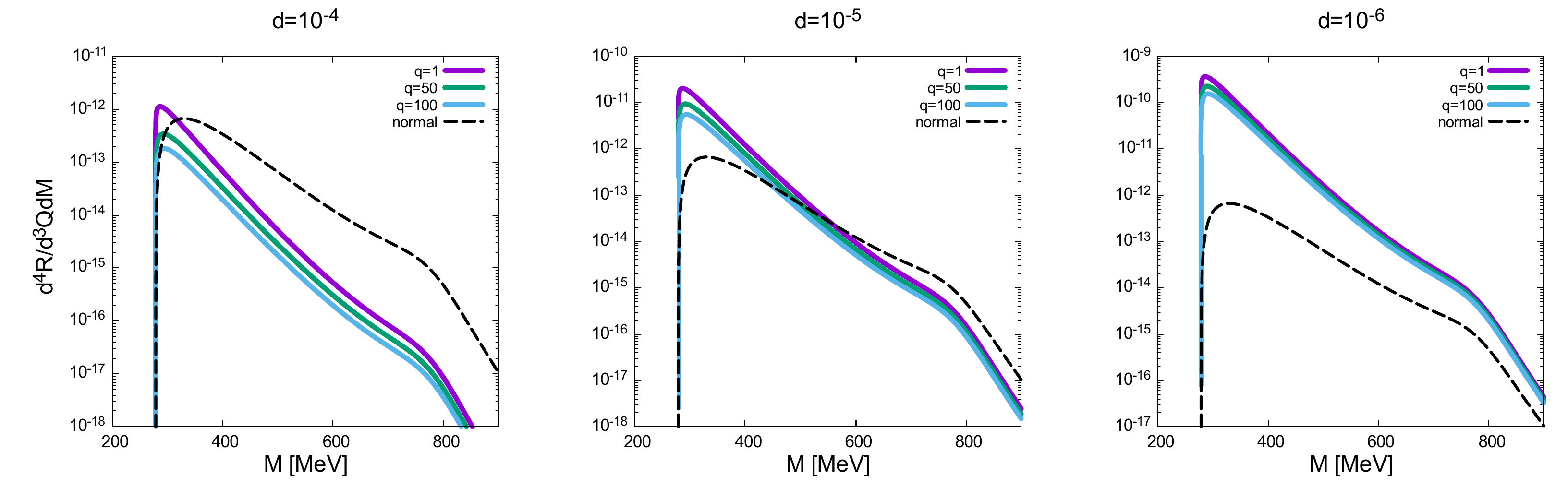}
\caption{These semi-logarithmic graphs show the electron-positron pair production rates by charged pion-pair annihilation 
as functions of the invariant mass $M$ at temperature $T=40\ \mathrm{MeV}$ based on the summation of Eq.\ \eqref{eq:PR_sum_replace}. 
From left to right, 
the parameter $d$ is fixed as $d=10^{-4},10^{-5},10^{-6}\ \mathrm{MeV}^{-2}$, respectively. 
The purple, green, and blue curves represent the values when $q=1, 50, 100\ \mathrm{MeV}$, respectively. The black dotted 
curves represent the production rates for normal dispersion 
$\omega^2=k^2+m_{\pi}^2$ in the homogeneous chiral condensed phase. }
\label{fig:PitoE_q_replace}
\end{figure*}

Now, let us set the parameters.
First, the temperature is set to $T=40\ \mathrm{MeV}$.
This is because the temperature at which the inhomogeneous chiral condensed phase appears is usually lower than about $50-100\  \mathrm{MeV}$ in previous studies by
using the effective models such as the NJL model \cite{Nakano:2005csa,Nickel:2009ipn}.
Although we do not specifically determine the value of baryon density (or quark chemical potential) here, we assume that its effect is included in the coefficients of the potential of Eq.\ \eqref{eq:V_GL}.
For $m_{\pi}$, $m_{\rho}$, $\Gamma_{\rho}$, and $m_{l}=m_{e}$, we adopt vacuum values: $m_{\pi}=139.6\ \mathrm{MeV},\ m_{\rho}=775\ \mathrm{MeV},\ \Gamma_{\rho}=149\ \mathrm{MeV},\ m_{e}=0.51\ \mathrm{MeV}$.
Of course, for these values we should take into account the effects of temperature and density, and inhomogeneous chiral condensations.
However, in this paper we ignore these and focus on the effect of the form of dispersion relations.
The coefficient $d$ in Eq.\ \eqref{eq:V_GL} and the wave number of the modulation $q$ are treated as free parameters.
These values should be identified experimentally or by specific models.
Although the Lagrangian density \eqref{eq:L_GL} with \eqref{eq:V_GL} is not valid unless $q$ is small, we also take somewhat larger values to see how the production rate changes with $q$.

Also, let us discuss how to add up the contributions from the cases of $f_+(\bm{k})$ and $f_-(\bm{k})$.
As can be seen from Eqs.\ \eqref{eq:Matrix_beta1_alpha2_k2},\eqref{eq:gamma+},\eqref{eq:gamma-}, the dispersion and mode corresponding to $f_+(\bm{k}-2\bm{q})$ are the same as those corresponding to $f_-(\bm{k}$), while the dispersion and mode corresponding to $f_-(\bm{k}+2\bm{q})$ are the same as those corresponding to $f_+(\bm{k})$.
In other words, the minimum point of dispersion of $f_+$ at $k\cos\theta/|\cos\theta| = -2q|\cos\theta|$ corresponds to the ground state of dispersion of $f_-$, while the minimum point of dispersion of $f_-$ at $k\cos\theta/|\cos\theta| = 2q|\cos\theta|$ corresponds to the ground state of dispersion of $f_+$.
Therefore, we assume that the part of $f_+$ in the range $k\cos\theta/|\cos\theta| \le -q|\cos\theta|$ is replaced by the part of $f_-$ in the range $k\cos\theta/|\cos\theta| \le q|\cos\theta|$, and the part of $f_-$ in the range $k\cos\theta/|\cos\theta| > q|\cos\theta|$ is replaced by the part of $f_+$ in the range $k\cos\theta/|\cos\theta| > -q|\cos\theta|$. Then we add up the contributions as
\begin{align}
\left. \frac{d^4 R}{d^3{Q} dM} \right|_{\bm{{Q}} = 0}
&= \left. \frac{d^4 R}{d^3{Q} dM} \right|_{\bm{Q} = 0,\omega^2=f_+ +m_{\pi}^2,k\frac{\cos\theta}{|\cos\theta|} \ge -q|\cos\theta|} \notag \\
&+\left. \frac{d^4 R}{d^3{Q} dM} \right|_{\bm{Q} = 0,\omega^2=f_- +m_{\pi}^2,k\frac{\cos\theta}{|\cos\theta|} < q|\cos\theta|}.
\label{eq:PR_sum_replace}
\end{align}

Figure \ref{fig:PitoE_q_replace} shows the electron-positron pair production rates in the inhomogeneous chiral condensed phase at some values of $d$ and $q$ based on the summation of Eq.\ \eqref{eq:PR_sum_replace}.
The black dotted curves represent the production rates for normal dispersion $\omega^2=k^2+m_{\pi}^2$ in the homogeneous chiral condensed phase. 
As can be seen from Eq.\ \eqref{eq:produtionrate_cal}, the dilepton production rate is inversely proportional to $d$.
At approximately $d\approx 10^{-5}\ \mathrm{MeV}^{-2}$, the production rates become close to those of the homogeneous chiral condensed phase.
If $d$ is larger or smaller, the production rates are clearly different.
Changing $q$ does not make much difference, but increasing $q$ will slightly reduce the production rates when $M \approx 2m_{\pi}$, that is, when $k$ is small. 
Basically, we can say that the overall slope is steeper than that in the case of normal dispersion $\omega^2=k^2+m_{\pi}^2$.
This is because when $k$ is large, the production rate is smaller due to the effect of the fourth-order term in Eq. \eqref{eq:dispersion_chargedpion}, and when $k$ is small, the production rate is larger due to the contribution of $\cos\theta\approx0$.

Also, we present numerical results in Appendix \ref{app:sum_all} where we sum the contributions without the above replacements.
In that case, the production rates diverge at $M=2m_{\pi}$.


\section{\label{sec:level6}summary}

In this paper, we have assumed a dual chiral density wave as an inhomogeneous chiral condensate, and derived the dispersion relations for the NG modes in the inhomogeneous chiral condensed phase from the low energy effective Lagrangian density expanded with respect to the order parameter based on $O(4)$ symmetry.
The derived dispersion relations are anisotropic and quadratic, and the corresponding modes appear as mixings of modes of broken symmetries due to inhomogeneous chiral condensation.

Also, we have evaluated the electron-positron production rates by charged pion-pair annihilation as functions of an invariant mass using the obtained dispersion relations.
Basically, the overall slope of the production rate versus the invariant mass is steeper than that
in the homogeneous chiral condensed phase.
Therefore,  the production rate may somewhat be enhanced around $M = 2m_{\pi}$, although it depends the value of $d$.
This may be a signature of the inhomogeneous chiral condensed phase.
In addition, a quantitative evaluation may be possible by determining the density, the coefficients in Eq.\ \eqref{eq:V_GL}, and the amplitude and the wave number in Eq.\ \eqref{eq:DCDW_gs}, which are treated as free parameters in this paper, using a specific model.

In appendix \ref{app:sum_all}, we present the numerical results without replacements as done in Sec.\ \ref{sec:level5} (Fig. \ref{fig:PitoE_q_noreplace}).
In that case, the production rate diverges at $M = 2m_{\pi}$ because of local minimum points of the dispersions $\omega^2=f_{\pm}(\bm{k})+m_{\pi}^2$ at $k\ne0$, which we ignored by the replacement in Sec.\ \ref{sec:level5}.
As discussed in Ref.\ \cite{Gale:1987drf}, divergences of dilepton production rate occur when the charged pion dispersion has any stationary points.
For example, similar divergence should occur when calculating using the moat spectrum $E=\sqrt{z\bm{k}^2+w\bm{k}^4+m^2}\ (z<0,w>0)$ as seen in Refs.\ \cite{Rennecke:2023pim, PR} ,although the threshold value will change.
Of course, in reality the divergence should be observed as a finite peak.

In this paper, we have adopted vacuum values as for pions and rho mesons, but they might change due to the effects of temperature and density.
In addition, they also may be affected by the effects of inhomogeneous chiral condensates in contrast to the ordinally homogeneous chiral condensate.
In particular, note that the constituent mass of quark is significantly reduced.
For example in Ref.\ \cite{Nakano:2003spc}, it is discontinuous decrease in the phase transition from homogeneous phase to inhomogeneous phase.
In addition, we have been discussing under small $\Delta$, which correspond to the constituent mass.
In this case, we may need to consider the decay of pion to $\bar{q}q$ because twice the quark mass is close to or smaller than $m_\pi = 139.6 \ \mathrm{MeV}$. 
However, it is thought that the pion mass will also be smaller in the DCDW phase.
In this case, the thresholds of dilepton production rates change depending on the pion mass.
Furthermore, it may be necessary to consider the effect of $\pi$-$N$ coupling at finite nucleon density. 
Besides, recalculations using the chiral $SU(3)$ theory should also be performed since strange quarks cannot be ignored at high density.
These are future problems.

In Ref.\ \cite{Lee:2015lpi}, using the same theory as in this paper, it is pointed out that the long range order of the inhomogeneous chiral condensed phase is destroyed by low energy fluctuations of the order parameter, and that this phase does not exist stably at finite temperature.
Instead, the possibility of a quasi-long-range order phase has been suggested.
Therefore, the inhomogeneous chiral condensed phase discussed in this paper may not be a stable phase.
However, this phase might also be stabilized in the presence of an external magnetic field \cite{TNK}.
Even if the inhomogeneous chiral condensed phase is metastable, it may still affect the dilepton pair production rate on processes of heavy ion collisions.
It would be also important to investigate precursor phenomena above the transition temperature and density, 
since achieving the low temperature and medium/high density region, in which the inhomogeneous chiral condensed phase may be realized theoretically, may be difficult 
in heavy ion collision experiment.


\begin{acknowledgments}

The authors acknowledge to the members of the Many-Body Theory Group in Kochi University.
After revising this paper by Referee's comments, we received a preprint \cite{Nsussinov:2024dpf} from Professor Robert D. Pisarski, in which the almost same issue is addressed. In Ref. \cite{Nsussinov:2024dpf}, it is pointed out that the higher-order terms of the $\pi\pi\gamma$ vertex are necessary with respect to recovering the gauge invariance and/or satisfying the Ward identity, and that they play a role in avoiding the divergence of the dilepton production rate at the threshold due to the minimum of the dispersion at non-zero momentum in the back-to-back case, while their argument is without the vector meson dominance and is not in the DCDW phase. Considering it in our study is a future problem. The authors would like to express their sincere thanks to Professor Robert D. Pisarski for providing us important information and for discussing with one of the authors (K. H.) directly.

\end{acknowledgments}


\appendix

\section{\label{app:SU2_O4}Chiral $SU(2)_{L} \times SU(2)_R$ theory and $O(4)$ theory}

First, we state the chiral $SU(2)_{L} \times SU(2)_{R}$ theory. 
Let $\psi_{L,R}$ be left-handed and right-handed two-flavor quark fields. 
The chiral $SU(2)$ transformations are given by $\psi_L \to \exp(-i\tau_{i}^{L}\theta_{i}^{L})\psi_L,\ \psi_R \to \exp(-i\tau_{j}^{R}\theta_{j}^{R})\psi_R \ (i,j=1,2,3)$, which has six generators $\tau_{1}^{L},\tau_{2}^{L},\tau_{3}^{L},\tau_{1}^{R},\tau_{2}^{R},\tau_{3}^{R}$.
Here, $\tau_{i}^{L}$ and $\tau_{j}^{R}$ are the Pauli matrices for left-handed and right-handed quarks, respectively.
If we set 
\begin{align}
\psi = 
\begin{pmatrix}
\psi_L \\
\psi_R \\
\end{pmatrix}
,\ 
\gamma_5 = 
\begin{pmatrix}
-1 & 0 \\
0 & 1 \\
\end{pmatrix} ,
\end{align}
the chiral $SU(2)$ transformations can be rewritten as vector transformations
\begin{align}
\psi \to U_V(\theta_{i}^{V})\psi \qquad (i=1,2,3), \\
U_V(\theta_{i}^{V}) = e^{-i\tau_i\theta_{i}^{V}} ,
\end{align}
and axial vector transformations \begin{align}
\psi \to U_A(\theta_{j}^{A})\psi \qquad (j=1,2,3), \\
U_A(\theta_{i}^{A}) = e^{-i\gamma_5\tau_j\theta_{j}^{A}} ,
\end{align}
where six generators are $\tau_{1},\tau_{2},\tau_{3},\gamma_5\tau_{1},\gamma_5\tau_{2}$ and $\gamma_5\tau_{3}$.
Now, considering the vacuum expectation values $\braket{\bar{\psi}\psi}$ and $\braket{\bar{\psi}i\gamma_5\vec{\tau}\psi}$ where $\vec{\tau} =(\tau_{1},\tau_{2},\tau_{3})$, $\braket{\bar{\psi}\psi}^2 +\braket{\bar{\psi}i\gamma_5\vec{\tau}\psi}^2$ is invariant under the chiral transformations.
Furthermore, $U_V(\theta_{1,2,3}^{V})$ and $U_A(\theta_{1,2,3}^{A})$ are the $\braket{\bar{\psi}i\gamma_5\tau_2\psi}$-$\braket{\bar{\psi}i\gamma_5\tau_3\psi}$, $\braket{\bar{\psi}i\gamma_5\tau_3\psi}$-$\braket{\bar{\psi}i\gamma_5\tau_1\psi}$, $\braket{\bar{\psi}i\gamma_5\tau_1\psi}$-$\braket{\bar{\psi}i\gamma_5\tau_2\psi}$, $\braket{\bar{\psi}\psi}$-$\braket{\bar{\psi}i\gamma_5\tau_1\psi}$, $\braket{\bar{\psi}\psi}$-$\braket{\bar{\psi}i\gamma_5\tau_2\psi}$ and $\braket{\bar{\psi}\psi}$-$\braket{\bar{\psi}i\gamma_5\tau_3\psi}$ rotations, respectively.

Next, we state the $O(4)$ theory. 
We introduce four-component field $\phi^{T} = (\sigma, \vec{\pi})$.
The $O(4)$ transformations are given by 
\begin{align}
&\psi \to U(\alpha_i)\psi, \ U(\alpha_i) = e^{-i\alpha_i L_{Vi}}\qquad (i=1,2,3), \\
&\psi \to U(\beta_j)\psi, \ U(\beta_j) = e^{-i\beta_j L_{Aj}}\qquad (j=1,2,3), 
\end{align}
which has six generators:
\begin{align}
L_{V1} = 
\begin{pmatrix}
0 & 0 & 0& 0 \\
0 & 0 & 0& 0 \\
0 & 0 & 0& -i \\
0 & 0 & i& 0 \\
\end{pmatrix}
,
L_{V2} = 
\begin{pmatrix}
0 & 0 & 0& 0 \\
0 & 0 & 0& i \\
0 & 0 & 0& 0 \\
0 & -i & 0& 0 \\
\end{pmatrix}
, \notag \\
L_{V3} = 
\begin{pmatrix}
0 & 0 & 0& 0 \\
0 & 0 & -i& 0 \\
0 & i & 0& 0 \\
0 & 0 & 0& 0 \\
\end{pmatrix}
,
L_{A1} = 
\begin{pmatrix}
0 & -i & 0& 0 \\
i & 0 & 0& 0 \\
0 & 0 & 0& 0 \\
0 & 0 & 0& 0 \\
\end{pmatrix}
, \notag \\
L_{A2} = 
\begin{pmatrix}
0 & 0 & -i& 0 \\
0 & 0 & 0& 0 \\
i & 0 & 0& 0 \\
0 & 0 & 0& 0 \\
\end{pmatrix}
,
L_{A3} = 
\begin{pmatrix}
0 & 0 & 0& -i \\
0 & 0 & 0& 0 \\
0 & 0 & 0& 0 \\
i & 0 & 0& 0 \\
\end{pmatrix} .
\end{align}
$U(\alpha_{1,2,3})$ and $U(\beta_{1,2,3})$ represent the $\pi_2$-$\pi_3, \pi_3$-$\pi_1, \pi_1$-$\pi_2, \sigma$-$\pi_1, \sigma$-$\pi_2$ and $\sigma$-$\pi_3$ rotations, respectively.

Comparing the two theories, $\sigma$ and $\vec{\pi}$ correspond to $\braket{\bar{\psi}\psi}$ and $\braket{\bar{\psi}i\gamma_5\vec{\tau}\psi}$, respectively. 
Also, $U(\alpha_{i})$ and $U(\beta_{j})$ correspond to vector rotation $U_V(\theta_{i}^{V})$ and pseudovector rotation $U_A(\theta_{j}^{A})$, respectively.


\section{\label{app:Mixed_Mode}Mixed modes}

In Appendix B we return to the discussion of the system with chiral $SU(2)_{L} \times SU(2)_R$ symmetry.
We adopt the NJL model as a concrete model with this symmetry. Lagrangian density is
\begin{align}
 \mathcal{L}_{\rm NJL}=\bar{\psi}i\gamma^\mu\partial_\mu\psi+G[(\bar{\psi}\psi)^2+(\bar{\psi}i\gamma_5\bm{\tau}\psi)^2] \ ,
\end{align}
where $G$ is a coupling constant.
The Noether currents $j^\mu_a$ and the Noether charge $Q^a$ are calculated as 
\begin{align}
&j_{A, a}^\mu=\frac{\partial {\cal L}_{\rm NJL}}{\partial (\partial_\mu \psi)}\delta_A^a\psi
=-{\bar \psi}\gamma^\mu\gamma_5\frac{\tau^a}{2}\psi\ , \notag \\
&j_{V, a}^\mu=\frac{\partial {\cal L}_{\rm NJL}}{\partial (\partial_\mu \psi)}\delta_V^a\psi
=-{\bar \psi}\gamma^\mu\frac{\tau^a}{2}\psi\ ,
\\
&Q_A^a=\int d^3{\bm x}\ j_{A,a}^{\mu=0}=-\int d^3{\bm x}\ {\bar \psi}\gamma^0\gamma_5\frac{\tau^a}{2}\psi\ , \notag\\
&Q_V^a=\int d^3{\bm x}\ j_{V,a}^{\mu=0}=-\int d^3{\bm x}\ {\bar \psi}\gamma^0\frac{\tau^a}{2}\psi\ , 
\end{align}
In the system with chiral $SU(2)_{L} \times SU(2)_R$ symmetry, the dual chiral density wave can be expressed as the following two condensates:
\begin{align}
\braket{\bar{\psi}\psi} = \Delta \cos (\bm{q}\cdot\bm{r}),\quad
\braket{\bar{\psi}i\gamma_5\tau^3\psi} = \Delta \sin (\bm{q}\cdot\bm{r}).
\label{eq:B_DCDW}
\end{align}
When these condensates have finite values, their spontaneous symmetry breaking is expressed by the following commutation relations:
\begin{align}
&\braket{[iQ_A^1, \bar{\psi}i\gamma_5\tau^1\psi]} = -\braket{\bar{\psi}\psi}, \notag \\
&\braket{[iQ_A^2, \bar{\psi}i\gamma_5\tau^2\psi]} = -\braket{\bar{\psi}\psi}, \notag \\
&\braket{[iQ_A^3, \bar{\psi}i\gamma_5\tau^3\psi]} = -\braket{\bar{\psi}\psi},
\label{eq:B_com_S1}
\end{align}
\begin{align}
&\braket{[iQ_V^1, \bar{\psi}i\gamma_5\tau^2\psi]} = \braket{\bar{\psi}i\gamma_5\tau^3\psi}, \notag \\
&\braket{[iQ_V^2, \bar{\psi}i\gamma_5\tau^1\psi]} = -\braket{\bar{\psi}i\gamma_5\tau^3\psi}, \notag \\
&\braket{[iQ_A^3, \bar{\psi}\psi]} = \braket{\bar{\psi}i\gamma_5\tau^3\psi}.
\label{eq:B_com_PS1}
\end{align}
Substituting Eq.\ \eqref{eq:B_DCDW} and making replacements $\bar{\psi}\psi \to \sigma,\quad  \bar{\psi}i\gamma_5\tau^a\psi \to \pi^a$, Eqs.\  \eqref{eq:B_com_S1} and \eqref{eq:B_com_PS1} yield
\begin{align}
&\braket{[iQ_A^1, \pi^1]} = -\Delta \cos (\bm{q}\cdot\bm{r}), \notag \\
&\braket{[iQ_A^2, \pi^2]} = -\Delta \cos (\bm{q}\cdot\bm{r}), \notag \\
&\braket{[iQ_A^3, \pi^3]} = -\Delta \cos (\bm{q}\cdot\bm{r}),
\label{eq:B_com_S2}
\end{align}
\begin{align}
&\braket{[iQ_V^1, \pi^2]} = \Delta \sin (\bm{q}\cdot\bm{r}), \notag \\
&\braket{[iQ_V^2, \pi^1]} = -\Delta \sin (\bm{q}\cdot\bm{r}), \notag \\
&\braket{[iQ_A^3, \sigma]} = \Delta \sin (\bm{q}\cdot\bm{r}).
\label{eq:B_com_PS2}
\end{align}
Combining Eqs.\  \eqref{eq:B_com_S2} and \eqref{eq:B_com_PS2}, we get the following commutation relations:
\begin{align}
\begin{cases}
\braket{[i(Q_A^1\cos(\bm{q}\cdot\bm{r})+ Q_V^2\sin (\bm{q}\cdot\bm{r})), \pi^1]} = -\Delta \\
\braket{[i(Q_A^1\sin (\bm{q}\cdot\bm{r})- Q_V^2\cos (\bm{q}\cdot\bm{r})), \pi^1]} = 0
\end{cases}
\\
\begin{cases}
\braket{[i(Q_A^2\cos(\bm{q}\cdot\bm{r})- Q_V^1\sin (\bm{q}\cdot\bm{r})), \pi^2]} = -\Delta \\
\braket{[i(Q_A^2\sin (\bm{q}\cdot\bm{r})+ Q_V^1\cos (\bm{q}\cdot\bm{r})), \pi^2]} = 0
\end{cases}
\\
\begin{cases}
\braket{[iQ_A^3, (\pi^3\cos (\bm{q}\cdot\bm{r}) - \sigma\sin (\bm{q}\cdot\bm{r}))]} = -\Delta \\
\braket{[iQ_A^3, (\pi^3\sin (\bm{q}\cdot\bm{r}) + \sigma\cos (\bm{q}\cdot\bm{r}))]} = 0
\end{cases}
\end{align}
From these commutation relations, the the Noether charges (generators) corresponding to the broken symmetries are
\begin{align}
&i(Q_A^1\cos(\bm{q}\cdot\bm{r})+ Q_V^2\sin (\bm{q}\cdot\bm{r})), \notag \\
&i(Q_A^2\cos(\bm{q}\cdot\bm{r})- Q_V^1\sin (\bm{q}\cdot\bm{r})), \notag \\
&iQ_A^3\ ,
\end{align}
and the NG bosons are 
\begin{align}
\pi^1,\quad
\pi^2,\quad
\pi^3\cos (\bm{q}\cdot\bm{r}) - \sigma\sin (\bm{q}\cdot\bm{r}).
\end{align}
From the correspondence in Appendix A, the NG bosons corresponding to the $\beta_1$-$\alpha_2$ mixing and $\beta_2$-$\alpha_1$ mixing modes in Sec.\ \ref{sec:level3} are thought to be $\pi_1$ and $\pi_2$.


\section{\label{app:Formulation_PR}Dilepton production rate}

Here, we derive expressions \eqref{eq:integration_theta} and \eqref{eq:produtionrate_theta} for the dilepton production rate from Eqs.\ \eqref{eq:PR_1}, \eqref{eq:iM} and \eqref{eq:form_factor}. 
First, let us calculate $|\mathcal{M}|^2$.
Now, the dispersion relation of the leptons is $E_{\pm}^2=\bm{p}_{\pm}^2+m_l^2$.
The current density due to the charged pions about the $\pi_+\pi_-\gamma$ vertex is
\begin{align}
j_{\mu}  = -i(\Phi_{\pi_{-}}\partial_{\mu}\Phi_{\pi_{+}} -\Phi_{\pi_{+}}\partial_{\mu}\Phi_{\pi_{-}}) ,
\end{align}
where $\Phi_{\pi_{\pm}}$ are the charged pion fields.
Considering $\Phi_{\pi_{\pm}}$ as a plane wave of the form $\exp(\mp ik^{\mu}x_{\mu})$, from the diagram of Eq.\ \eqref{eq:iM}, we obtain
\begin{align}
|\mathcal{M}|^2 
&= \frac{e^4}{q^4}  (k_1 -k_2)_{\mu} (k_1 -k_2)_{\nu} \notag \\
   &\quad \times \mathrm{Tr}((\Slash{p}_{-} +m_l)\gamma^{\mu}(\Slash{p}_{+} -m_l)\gamma^{\nu}) .
\label{eq:AC_M2}
\end{align}

Substituting Eq.\ \eqref{eq:AC_M2} into Eq.\eqref{eq:PR_1} and performing $\int d^3k_2$ and $\int d^3p_-$, Eq.\ \eqref{eq:PR_1} becomes 
\begin{align}
\left. \frac{d^4 R}{d^3{Q} dM} \right|_{\bm{{Q}} = 0}
&= \frac{8\alpha^2}{(2\pi)^6 M^4} |F_{\pi}(M^2)|^2 \int\frac{d^3k_1}{2\bar{\omega}} f_1 f_2 \notag \\
   &\quad \times \int\frac{d^3p_{+}}{E_{+}^2} \left[ (1-\cos^2 \varphi)\bm{k}_1^2\bm{p}_{+}^2 +m_l^2\bm{k}_1^2) \right] \notag \\
   &\quad \times \frac{1}{2}\delta \left(\frac{M}{2}-E_{+} \right) \delta(M -2\bar{\omega}) ,
\label{eq:AC_PR_intp}
\end{align}
where $\varphi$ is the angle between $\bm{p}_+$ and $\bm{k}_1$ and $\bar{\omega} \equiv ({\omega_1+\omega_2})/{2} = \omega_1=\omega_2$.
Let us replace $\int d^3p_+$ with the polar integral: 
$\int d^3p_+=\int_0^{\infty} \bm{p}_+^2 d|\bm{p}_+| \int_0^{\pi} \sin\varphi d\varphi \int_{-\pi}^{\pi} d\varphi'$.
Here, we identify the angle $\varphi$ between $\bm{p}_+$ and $\bm{k}_1$ as the integral variable $\varphi$.
Because $\int_0^{\infty} \bm{p}_+^2 d|\bm{p}_+|=\int_0^{\infty} E_+^2 dE_+$, Eq.\ \eqref{eq:AC_PR_intp} yields
\begin{align}
\left. \frac{d^4 R}{d^3{Q} dM} \right|_{\bm{{Q}} = 0}
&= \frac{4\alpha^2}{3(2\pi)^5} \frac{M^2+2m_l^2}{M^4}  |F_{\pi}(M^2)|^2  \notag \\
   &\quad \times \int\frac{d^3k_1}{2\bar{\omega}} f_1 f_2 \bm{k}_1^2 \delta(M -2\bar{\omega}) .
\end{align}
As in the case of $\int d^3p_+$, let us replace $\int d^3k_1$ with the polar integral: 
$\int d^3k_1=\int_0^{\infty} \bm{k}_1^2 d|\bm{k}_1| \int_0^{\pi} \sin\theta d\theta \int_{-\pi}^{\pi} d\theta'$.
Here, we identify the variable $\theta$ of $\bar{\omega}(|\bm{k}_1|,\theta)$ as the integral variable $\theta$.
\begin{align}
\left. \frac{d^4 R}{d^3{Q} dM} \right|_{\bm{{Q}} = 0}
&= \int_0^{\pi} d\theta \frac{4\alpha^2}{3(2\pi)^4} \frac{M^2 +2m_l^2}{M^4} |F_{\pi}(M^2)|^2  \notag \\
   &\quad \times \int_0^{\infty} d|\bm{k}_1|
    \frac{\bm{k}_1^4 \sin\theta}{2\bar{\omega}} f_1 f_2 \delta(M -2\bar{\omega}) .
\end{align}
\begin{figure*}[tbp]
 \centering
 \includegraphics[keepaspectratio, height=5cm]{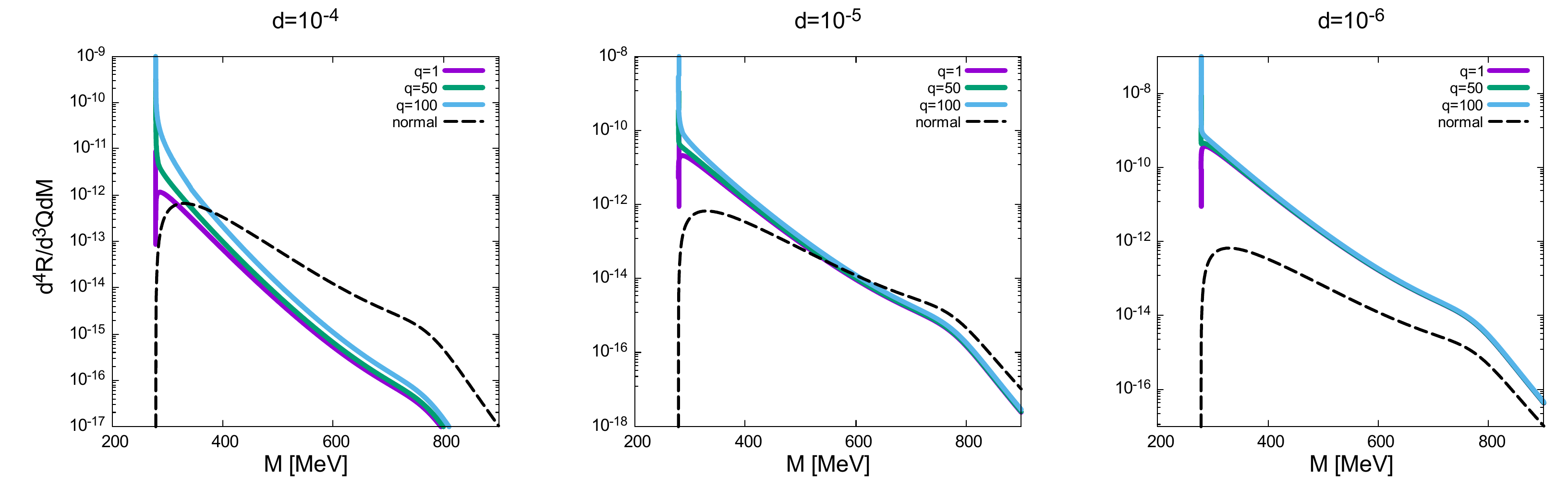}
 \caption{These figures are the same as FIG.\ \ref{fig:PitoE_q_replace}, except that they are based on the summation of Eq.\ \eqref{eq:PR_sum_noreplace}.
For any parameter sets of $d$ and $q$, the production rate diverges at $M = 2m_{\pi} \approx 279\ \mathrm{MeV}$. }
\label{fig:PitoE_q_noreplace}
\end{figure*}
Now we convert $\int d|\bm{k}_1|$ to $\int d\bar{\omega}$.
The Jacobian is $\left| {d\bar{\omega}}/{dk} \right|^{-1}$, and the $\int d\bar{\omega}$ becomes the sum over all that satisfies $2\bar{\omega}(|\bm{k}_1|,\theta) =M$ because of $\delta(M -2\bar{\omega}) $.
That is,
\begin{align}
&\int_0^{\infty} d|\bm{k}_1| F(|\bm{k}_1|,\bar{\omega}) \delta(M -2\bar{\omega}) \notag \\
&\quad \to \sum_{j} F(k^{(j)}, M/2) \left| \frac{d\bar{\omega}}{dk} \right|_{k=k^{(j)}, 2\bar{\omega} = M}^{-1} ,
\end{align}
where $k^{(j)}$ is one of the solutions of $|\bm{k}_1|$ that satisfies $2\bar{\omega}(|\bm{k}_1|,\theta) =M$.
Therefore, we get expressions \eqref{eq:integration_theta} and \eqref{eq:produtionrate_theta}.


\section{\label{app:sum_all}Numerical results without replacements}

In Sec.\ \ref{sec:level5}, we present the numerical results based on the summation of Eq.\ \eqref{eq:PR_sum_replace}.
Namely, they are the results calculated under replacing the part of $f_+$ in the range $k\cos\theta/|\cos\theta| \le -q|\cos\theta|$ with the the part of $f_-$ in the range $k\cos\theta/|\cos\theta| \le q|\cos\theta|$, and $f_-$ in the range $k\cos\theta/|\cos\theta| > q|\cos\theta|$ with the part of $f_+$ in the range $k\cos\theta/|\cos\theta| > -q|\cos\theta|$.

In this appendix, we calculate without the replacements stated above, instead multiplying by a factor of $1/2$.
Therefore, we add up the contributions as
\begin{align}
\left. \frac{d^4 R}{d^3{Q} dM} \right|_{\bm{{Q}} = 0}
&= \frac{1}{2} \left. \frac{d^4 R}{d^3{Q} dM} \right|_{\bm{Q} = 0,\omega^2=f_+ +m_{\pi}^2} \notag \\
&+\frac{1}{2} \left. \frac{d^4 R}{d^3{Q} dM} \right|_{\bm{Q} = 0,\omega^2=f_- +m_{\pi}^2}\ .
\label{eq:PR_sum_noreplace}
\end{align}

Figure \ref{fig:PitoE_q_noreplace} shows the electron-positron pair production rates of the inhomogeneous chiral condensed phase at some values of $d$ and $q$ based on the summation of Eq.\ \eqref{eq:PR_sum_noreplace}.
The black dotted curves are the same as in Fig.\ \ref{fig:PitoE_q_replace}.
Just as the case of Eq.\ \eqref{eq:PR_sum_replace}, the production rate is inversely proportional to $d$.
In contrast to Fig.\ \ref{fig:PitoE_q_replace},  increasing $q$ will slightly increase the production rates when  $k$ is small.
The most notable point is that the production rates diverge when $M = 2m_{\pi}$.
Such a divergence occurs when $\left| {d\bar{\omega}}/{dk} \right|$ in Eq.\ \eqref{eq:produtionrate_theta} equals $0$ 
at $k\ne0$.
As can be seen from Fig.\ \ref{fig:dispersion_f+_f-}, the dispersion of $f_+$ has a local minimum point at $k=-2q|\cos\theta|$ and $\omega_{i}=m_{\pi}$, i.e., $M=2m_{\pi}$, while the dispersion of $f_-$ has a local minimum point at $k=2q|\cos\theta|$ and $\omega_{i}=m_{\pi}$.
These point is the cause of the divergence of the production rate, and conversely the existence of the divergence may be seen as a signature of such dispersion.
Also, as in the case of Fig.\ \ref{fig:PitoE_q_replace}, the overall slope is steeper than that in the case of normal dispersion.


\end{document}